\documentclass[aps,prl,twocolumn,groupedaddress]{revtex4-1}
\usepackage{graphicx}
\usepackage{amsmath}
\usepackage{textcomp}
\usepackage{mathrsfs}

\usepackage[all]{xy}

\begin{document}

\title{From waves to bullets: testing Feynman's idea on the two slit experiment}

\author{Marco Ornigotti$^1$}
\email{marco.ornigotti@uni-jena.de}
\author{Andrea Aiello$^{2,3}$}
\affiliation{$^1$ Institute of Applied Physics, Friedrich-Schiller-Universit\"at Jena, Max-Wien-Platz 1, D-07743 Jena, Germany}
\affiliation{$^2$ Max Planck Institute for the Science of Light, G$\ddot{u}$nther-Scharowsky-Strasse 1/Bau24, 91058 Erlangen, Germany} 
\affiliation{$^3$ Institute for Optics, Information and Photonics, University of Erlangen-Nuernberg, Staudtstrasse 7/B2, 91058 Erlangen, Germany}

%\homepage[]{Your web page}
%\thanks{}

\date{\today}

\begin{abstract}
We test the validity of Feynman's idea that a two-slit experiment performed with classical objects (bullets) does not produce observable interference fringes on the detection screen because the Compton's wavelength of the bullets is so tiny, that no real detector could resolve individual interference fringes, thus producing only an average signal which is the observed smooth curve. To test this idea, we study the two-slit experiment in two different situations using light to simulate both wave-like  and particle-like bullets. In the first case, we consider coherent light with short wavelengths and in the second case incoherent light with not-so-short wavelength. While in the former case (simulating Feynman's wave-like bullets) the interference fringes are so dense that they cannot be resolved by a detector, therefore resulting in an averaged smooth signal, in the latter case (simulating Feynman's particle-like bullets), although the detector is fully capable of discriminating each fringe, the observed  classical smooth pattern limit is produced because of the lack of spatial coherence of the impinging field.
\end{abstract}

\maketitle
\section{I. Introduction}
In the first chapter of the third volume (Quantum Mechanics) of the celebrated series of \emph{The Feynman lectures on physics}, Richard Feynman gave a deep and detailed account of the paradigmatic ``two-slit experiment'', key to the foundation of quantum mechanics \cite{feynman}. In Feynman's words, studying this experiment
\begin{quote}
``[...] we shall tackle immediately the basic element of the mysterious behavior in its most strange form. We choose to examine a phenomenon which is impossible, absolutely impossible, to explain in any classical way, and which has in it the heart of quantum mechanics. In reality, it contains the \emph{only} mystery. We cannot make the mystery go away by ``explaining'' how it works. We will just tell you how it works. In telling you how it works we will have told you about the basic peculiarities of all quantum mechanics.''
\end{quote}
Then, Feynman proceeded illustrating, in its unique and whimsical style,  this experiment operated first with classical particles (bullets), second with classical waves (water waves) and, finally, with \emph{quantum} particles (electrons). While in the last two cases an interference pattern is observed on the detection screen, in the first case the  distribution of the bullets on the detector shows no interference. After a thorough and detailed analysis of these three examples, Feynman concluded that the motion of all matter must be described in terms of waves. However, he argued, 
\begin{quote}
``If the motion of all matter--as well as electrons--must be described in terms of waves, what about the bullets in our first experiment? Why didn't we see an interference pattern there? It turns out that for the bullets the wavelengths were so tiny that the interference patterns became very fine. So fine, in fact, that with any detector of finite size one could not distinguish the separate maxima and minima. What we saw was only a kind of average, which is the classical curve.''
\end{quote}
Does this last sentence sound meaningful? As a matter of fact, it is hard to swallow  that, according to Feynman, even billiard balls would produce an interference pattern, although so dense that it cannot be seen. Rather, it is easier to think that in such a case interference fringes would not be produced at all. Which viewpoint is the correct one?

In this paper we aim at testing the validity of Feynman's belief. Namely, we try to understand to what extent is true that the classical curve obtained in the experiment performed with bullets may be thought as resulting from the interference of waves with very short wavelength.
To accomplish our goal, we study the two-slit experiment performed with light of two different kinds: \emph{a}) coherent with short wavelength and \emph{b}) incoherent with not-so-short wavelength. Here, coherent and incoherent refer to the \emph{spatial} distribution of light (in this work we consider only monochromatic waves which are, by definition, temporally coherent). Moreover, ``short'' and ``not-so-short'' wavelengths are relative to the spatial resolution of the detector: in the first case the wiggles of the interference pattern are so dense that  cannot be resolved by the detector which, unavoidably, averages over several fringes. Vice versa, in the second case the interference pattern, when present, would be sparse and the detector could discriminate each fringe. As a result of our analysis, we obtain in both cases \emph{a}) and \emph{b}) the same ``classical curve'', thus vindicating and validating Feynman's belief.

This work is organized as follows: in Sect. II, we briefly review the problem of the diffraction of an electric field from a two slit aperture, and we will calculate the intensity distribution on a screen in the far field. In Sect. III, we concentrate our attention in proving how the averaging effect of a detector placed on the screen allows to obtain the classical (particle-like) regime as a limiting case of the quantum (wave-like) one. Section IV is instead devoted to approach the problem of the classical limit of a quantum particle by means od decoherence. Finally, conclusions are drawn in Sect. V.

\section{II. TWO SLIT EXPERIMENT WITH PLANE WAVES}
\begin{figure}[t!]
\begin{center}
\includegraphics[width=0.5\textwidth]{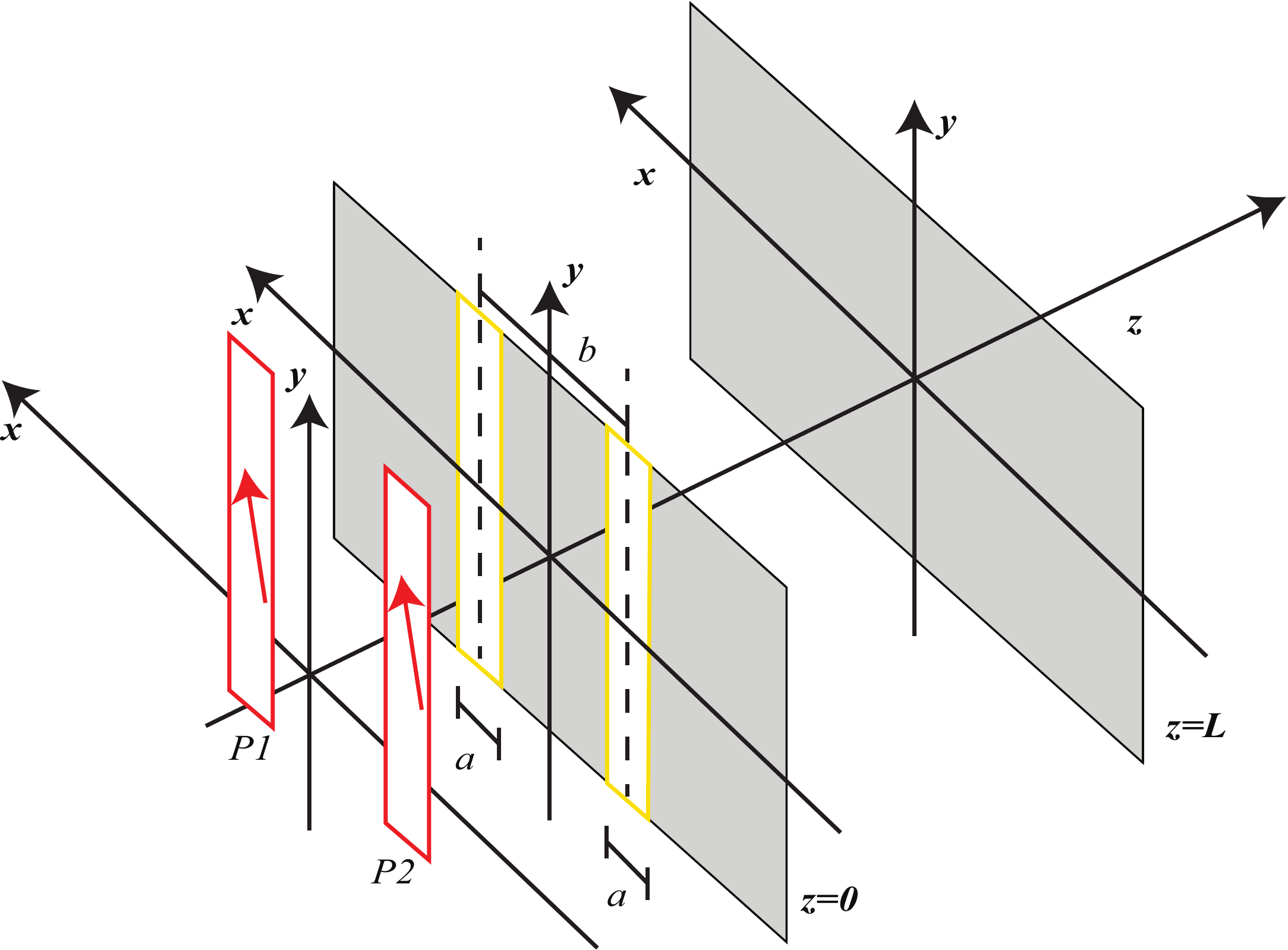}
\caption{Geometry of the two slit experiment. The two slits (depicted with yellow borders in the plane $z=0$) are separated by a distance $b$ and have length $a$ in the $x$-direction and they extend towards infinite into the $y$-direction. The detector screen is placed at a distance $L$ from the slits. The two polarizers $P1$ and $P2$ can be put before the slits and give, if needed, different polarization to the two beams.} 
\label{figura1}
\end{center}
\end{figure}

Consider a monochromatic plane wave of wavelength $\lambda$ and angular frequency $\omega=2\pi c/\lambda$, propagating along the $z$ direction that firstly impinges on a screen in $z=0$, where two slit apertures of length $a$ along the $x$-direction (and infinitely extended along the $y$-direction) and displaced along the transversal $x$-axis of a quantity $b$ are made. The two slits then generate two different waves that propagate and diffract from the slit plane $z=0$ to the screen plane at $z=L$, where a detector sensible to light intensity is present to reveal the transversal intensity pattern of the diffracted light. In order to exploit the polarization degree of freedom of light for studying how the coherence properties of the two beams affect the interference pattern on the detector screen, we imagine to put two polarizers right before the slit plane $z=0$, as shown in Fig. \ref{figura1}. 
By assuming that the two polarizers of Fig. \ref{figura1} assign in general	 two different polarizations $\hat{\mathbf{p}}_1=\cos\theta_1\hat{\mathbf{x}}+\sin\theta_1\hat{\mathbf{y}}$ and $\hat{\mathbf{p}}_2=\cos\theta_2\hat{\mathbf{x}}+\sin\theta_2\hat{\mathbf{y}}$ ($\theta_k$ ($k\in\{1,2\}$) is the polarization angle in the transverse $xy$ plane for the polarizer $P_k$) to the electric fields passing through the two, the total electric field of the beam in the slit plane can be written as:\\
\begin{eqnarray}\label{E_init}
\mathbf{E}_{in}(x,0,t) & = & E_0e^{-i\omega t}\Big[ \mathrm{rect}\Big(\frac{x-b/2}{a}\Big)\hat{\mathbf{p}}_1\nonumber \\
	& + &  \mathrm{rect}\Big(\frac{x+b/2}{a}\Big)\hat{\mathbf{p}}_2\Big]\nonumber\\
	&\equiv& E_1(x,z,t)\hat{\mathbf{p}}_1+E_2(x,z,t)\hat{\mathbf{p}}_2,
\end{eqnarray}
where $ \mathrm{rect}(x/a)$ is the rectangle function, that has the value one inside the interval $[x-a/2;x+a/2]$ and it is zero elsewhere. Since the temporal structure of the electric field does not play any role in this model, we can, without any loss of generality, disregard it for the rest of the paper. 

In order to study the fringe pattern generated by the interference of the two waves diffracted by the slits, we exploit Fraunhofer scalar diffraction theory, i.e., we put the detector screen at a distance $z\rightarrow\infty$ with respect to the slit plane at $z=0$. From diffraction theory \cite{goodman}, it is known that this limit is achieved when $L\geq 4a^2/\lambda$. Then the field distribution on the detector screen can be easily obtained by applying the Fraunhofer propagator to the initial field distribution as follows:\\
\begin{eqnarray}\label{fraunhofer}
\mathbf{E}(x,y,L)&=&\frac{e^{ik(L+\frac{x^2}{2L})}}{i\lambda L}\int d\xi\Big[E_1(\xi,0)\hat{\mathbf{p}}_1\nonumber\\
&+&E_2(\xi,0)\hat{\mathbf{p}}_2\Big]e^{-\frac{ik}{z}(x\xi)}
\end{eqnarray}
The reader may note that Eq. \eqref{fraunhofer} represents the Fourier transform of the field distribution right after the slit plane $z=0$ that has been propagated to the detector plane $z=L$. In order to calculate the previous integral, it is useful to remember the Fourier transform of the rectangle function. According to Ref. \cite{goodman} we have:

\begin{equation}\label{FTrect}
\int_{-\infty}^{+\infty}  \mathrm{rect}\Big(\frac{t-t_0}{a}\Big)e^{i\tau t}dt=\frac{|a|}{\sqrt{2\pi}}e^{it_0\tau}\mathrm{sinc}\Big(\frac{a\tau}{2}\Big),
\end{equation}
where $\mathrm{sinc}(x)=\sin(x)/x$ is the cardinal sine function. 

Then, by means of the superposition principle, the field distribution on the detector screen can be calculated as the superposition of the field diffracted from one of the slits plus the field diffracted from the other slit as $\mathbf{E}(x,L)=\mathbf{E}_1(x,L)+\mathbf{E}_2(x,L)$, where the two fields $\mathbf{E}_1(x,y,L)$ and $\mathbf{E}_2(x,y,L)$ can be calculated by substituting their expression as defined in Eq.\eqref{E_init} into Eq. \eqref{fraunhofer} and by using Eq. \eqref{FTrect} to calculate the integrals. After some simple algebra we obtain:
\begin{subequations}\label{E_screen}
\begin{equation}
\mathbf{E}_1(x,L) = Ce^{\frac{ik}{2L}x^2}\mathrm{sinc} \Big(\frac{ax}{2}\Big)e^{i(\frac{kb}{2L})x}\hat{\mathbf{p}}_1,
\end{equation}
\begin{equation}
\mathbf{E}_2(x,L)=Ce^{\frac{ik}{2L}x^2}\mathrm{\mathrm{sinc}}\Big(\frac{ax}{2}\Big)e^{-i(\frac{kb}{2L})x}\hat{\mathbf{p}}_2,
\end{equation}
\end{subequations}
where $C=-i(E_0 a^2)/(2\pi\lambda L)$ is an inessential proportionality constant.	

The intensity is thus simply proportional to the square modulus of the total field on the detector screen, i.e., $I(x,L)\propto |\mathbf{E}_1(x,L)+\mathbf{E}_2(x,L)|^2$, whose expression is the following:\\
\begin{equation}\label{intensity}
I(x,L) = 2|C|^2\mathrm{sinc}\Big(\frac{ax}{2}\Big)^2
\Big [1 + (\hat{\mathbf{p}_1}\cdot\hat{\mathbf{p}}_2)\cos\Big(\frac{kb}{L}x\Big)\Big],
\end{equation}
where we have used the fact that $|\hat{\mathbf{p}}_1|=|\hat{\mathbf{p}}_2|=1$, since they are unit vectors.

In section III we will use this formula by assuming that both beams have the same polarization when exiting from the slits and concentrate ourselves on the modeling of the detector that collects light on the detector screen, showing how the \emph{classical} (particle-like) regime is obtained as a limiting case of the \emph{quantum} (i.e., wave-like) behavior when the wavelength of the light that impinges on the detector goes to zero, i.e.,  it is much smaller than the detector characteristic length.  This will show that heavy-mass objects do not show interference because their wavelength is too tiny to be detected, i.e.,  they behave like \emph{classical} particles \cite{pakstakos}.

In section IV, instead, we will assume that the two polarizers $P_1$ and $P_2$ will introduce random polarization on the two beams, and we will analyze the effect of the decoherence of the two beams on the interference fringes of the intensity distribution \eqref{intensity}. This will show instead that the lack of quantum effects in a classical particle can be viewed as a decoherence effect among all the wave functions of the atoms that compose the particle itself \cite{peres_dec}.

\section{III. THE EFFECT OF THE MASS OF PARTICLES}
On his famous lecture on physics, Feynman discusses the absence of interference fringes for classical particles (i.e., bullets) with the following argument \cite{feynman}:

 \begin{quote}
 ``It turns out that for the bullets the wavelengths were so tiny that the interference patterns became very fine. So fine, in fact, that with any detector of finite size one could not distinguish the separate maxima and minima. What we saw was only a kind of average, which is a classical curve. "
 \end{quote}

In this section we want to prove Feynman's argument by considering a simple model for the detector placed on the observation screen and investigate what happens in the limit in which the light that impinges on the detector has a wavelength $\lambda\rightarrow 0$, a situation that reproduces the quantum behavior of a classical object. We then expect that in this limit, what we see on the detector will be the classical probability distribution to find the particle as the bare sum of the probability that the particle is passed through the left \emph{or} right slit, i.e.,  we expect to see on the detector plane that $I(x,y,L)\propto I_1(x,y,L)+I_2(x,y,L)$, with no interference at all.
We also note that this limit ($\lambda\rightarrow 0$) is analog to the limit presented in Ref. \cite{pakstakos}, where the author uses gaussian wavepackets (that are minimal uncertainty states, i.e.,  the more classical among the quantum states of a microscopic system) to describe either a microscopic and a macroscopic particle, underlining how the mass plays a central role for determining the level of ``\emph{quantumness}" of the particle itself. In particular, in the limiting case when $m\rightarrow\infty$ no more quantum features can be seen.

This comes immediately from our situation if the reader will recall the De Broglie relation that relates wavelength and momentum, i.e.,  $\lambda=h/(mv)$. From this formula one can immediately see that in the limit in which the mass of the particle is very high, its wavelength is very tiny: this is precisely the Feynman argument.

Before investigating in detail this limit, let us first set up a suitable model for the detector placed on the screen plane ($z=L$).\\

\subsection{A. Model for the Detector}
A very simple abstraction of a photo-detector is a two-dimensional object that can convert the incoming light into an electric signal proportional to the intensity of the detected light. Given this building block, we can imagine to build bigger detectors by simply juxtaposing these smaller objects (in the same way as a CCD camera is just a finite-size two dimensional arrangement of single detectors). In particular, we can imagine to fill the screen plane ($z=L$) of our experiments with a two-dimensional arrangement of such light detectors. In this case, we will have as output signal a collection of electric signals (each coming from a single detector on the screen), that will reproduce with a certain fidelity the intensity of the light field impinging on the screen at $z=L$. Although this model is already simple and satisfactory, we can simplify it even more by noting that, since the two slits are infinitely extended along the $y$-direction, the system possesses translational symmetry along that direction. This gives us the possibility to focus only on a region of screen of constant height (say, for the sake of simplicity, $y=0$), and acquiring only the field distribution along the $x$-direction. Thanks to this simplification, we can reduce our model for the detector screen from a two-dimensional arrangement of photo-detectors to a one-dimensional array of  such detectors centered around the $z$-axis, giving us the possibility to reduce the acquired signal from a two-dimensional set of data to a one-dimensional string. In order to give a better characterization to this model, we assume to consider an array of $N$ identical square detectors, each of length $L_{det}$.

After having established a valid (and physically meaningful) model for the detector, we now must turn our attention to the output signal, trying to find a valid relation that links the output of the detector (electric signal) to its input (light signal). Without any loss of generality (and according to standard detection models) we can imagine that the detector acquires the light signal impinging on its surface, thus producing an output electric signal that is proportional to the incident intensity averaged over the detector area. For the case of our $N$-detector array, the acquired intensity can be thus written as
\begin{equation}\label{acquired}
I_{acquired}=\sum_{k=1}^{N}I_k^{det}=\sum_{k=1}^{N}\frac{1}{L_{det}}\int_{-d/2}^{d/2} I(x_k,L)dx_k,
\end{equation}
where $x_k$ is the coordinate of a cartesian reference frame centered on the $k^{th}$ detector, so that for each detector the integration is carried out only for the portion of light intensity that impinges directly on it. Note that, by virtue of translational symmetry along the $y$-direction, we can consider our detectors as effectively one-dimensional objects, rather than physically two-dimensional.

The continuous light pattern on the screen is then transformed into a discrete set of counts that can be easily visualized in an histogram, where each bar corresponds to a single detector and the height of that bar is $I^{det}_k$. The resolution of this apparatus depends on two parameters. We can define a global resolution, that depends on the number of detectors we use for sampling the screen. If we use too less detectors, we will sample only a small part of the screen, making the interference pattern hard to recognize, while if we use a number of detectors sufficient for an optimal sampling of the screen, then the interference pattern on the screen will be reproduced with high fidelity. Without loss of generality, we can assume to be in this latter situation, where the $N$ detectors cover all the light dynamics along the $x$-axis.

Most interesting for our purposes is the second kind of resolution, that we can call \emph{local} resolution, and is given by the characteristic length $L_{det}$ of the single detector compared to the spatial period $\Lambda$ of the interference pattern we want to recognize. If $L_{det}<\Lambda$, a single oscillation (fringe) will be recorded by more than one single detector, thus making the maximum and the minimum of the fringe clearly visible. On the other hand, if $L_{det}\gg\Lambda$, i.e., if the fringes are oscillating very rapidly, a great number of fringes will be acquired by the same detector. This means that the output signal will be the average of a rapidly oscillating incoming signal over the detector length, namely a not oscillating quantity anymore. This will give as a result that, while the incoming light signal presents rapid oscillations, the acquired signal does not oscillate anymore. Then, for optimizing the acquired signal, a detector length smaller than the spatial period $\Lambda$ of the fringe pattern must be chosen. 

The reader should be aware of the fact that since the detector is a physical object, once this length is determined, there is no possibility anymore to dynamically change it. This means that if we chose $L_{det}$ to match a specific spatial period $\Lambda$, then we have implicitly selected a reference wavelength $\bar{\lambda}$ (to which $\Lambda$ is related) for which the acquiring operation is optimized. If we now change the wavelength of the impinging radiation, the local resolution will also change, progressively degradating if $\lambda\rightarrow 0$.

Among all the possible choices, we adopt a detector whose length is exactly half of the fringe period at a fixed wavelength, i.e.,  $L_{det}=\Lambda/2$, where $\Lambda=\bar{\lambda}L/(2b)$ is the spatial periodicity of the interference term in Eq. \eqref{intensity} and $\bar{\lambda}$ represents the chosen reference wavelength. The value of $\Lambda$ can be easily calculated by applying the definition of periodic functions, namely $\cos(\kappa x)=\cos(\kappa x+2\pi)$, thus obtaining $\Lambda=2\pi/\kappa$. In the case of Eq. \eqref{intensity} we have $\kappa=(kb)/L$ and therefore we obtain $\Lambda=\lambda L/(2b)$.

In order to compare the real interference pattern (the one that arrives on the screen and it is described by Eq. \eqref{intensity}) with the detected pattern (Eq. \eqref{acquired}), we need to simulate the action of the detectors numerically. A very simple way to do it is to write a routine that needs to receive as an input the analytic form of the interference pattern (namely, Eq. \eqref{intensity}), the number of detectors used, and the characteristic length of each detector. Once these data are available, the routine should basically cycle through each of the detector producing, for each of them, the averaged signal $I_k^{det}$, according to some pre-determined averaging algorithm, that in our case is given by Eq. \eqref{acquired} itself.

\subsection{B. Classical Limit for Heavy-Mass Particles}
We now present the results of the two slit experiment when on the screen at $z=L$ is placed a detector like the one described above, and for the case of two parallel polarized beams ($\hat{\mathbf{p}}_1\cdot\hat{\mathbf{p}}_2=1$). The parameters of the slit are assumed to be $a=1$ mm and $b=3$ mm. We use as a reference wavelength $\bar{\lambda}=800$ nm. The position of the detection screen is chosen to be as far as needed for exploiting Fraunhofer regime, i.e.,  $L=4a^2/\lambda$.

With these parameters,  the length of a single detector will be $L_{det}=\Lambda/2=\bar{\lambda}L/(2b)=65\mu$m. In order to properly cover all the dynamics of the light pattern in the $x$-direction (whose complete extension can be determined by plotting Eq. \eqref{intensity} and set to be $L_{tot}=24$ mm), we choose to use $N=L_{tot}/L_{det}=370$ detectors.

As a figure of merit for comparing the different cases, for each presented case (including the one including the polarization decoherence that will be treated in the next section) we will compute the fringe visibility with the help of the standard formula \cite{mandelWolf}\\
\begin{equation}
V=\frac{I_{max}-I_{min}}{I_{max}+I_{min}}.
\end{equation}

\begin{figure}[t!]
\begin{center}
\includegraphics[width=0.5\textwidth]{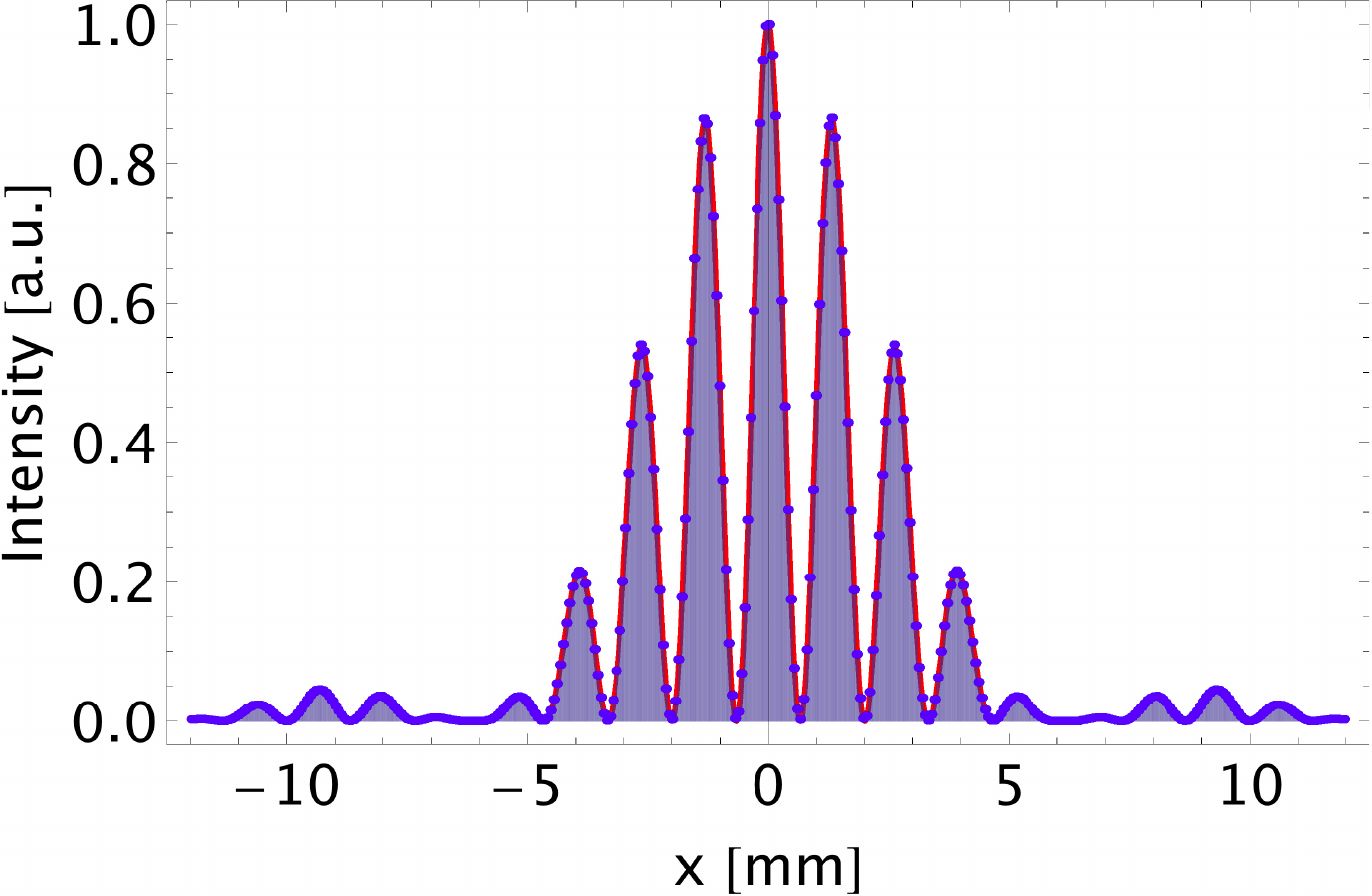}
\caption{(color online) Analytic interference pattern (red solid line) and output signal from the detectors (blue vertical bars) corresponding to the reference wavelength of $\lambda=800$ nm.  The fringe visibility, calculated by means of the intensity values detected by the detector, is $V=0.994$. As can be seen, in this case the detector is able to completely reconstruct the interference profile of the two beams.}
\label{figura2}
\end{center}
\end{figure}

\begin{figure}[t!]
\begin{center}
\includegraphics[width=0.5\textwidth]{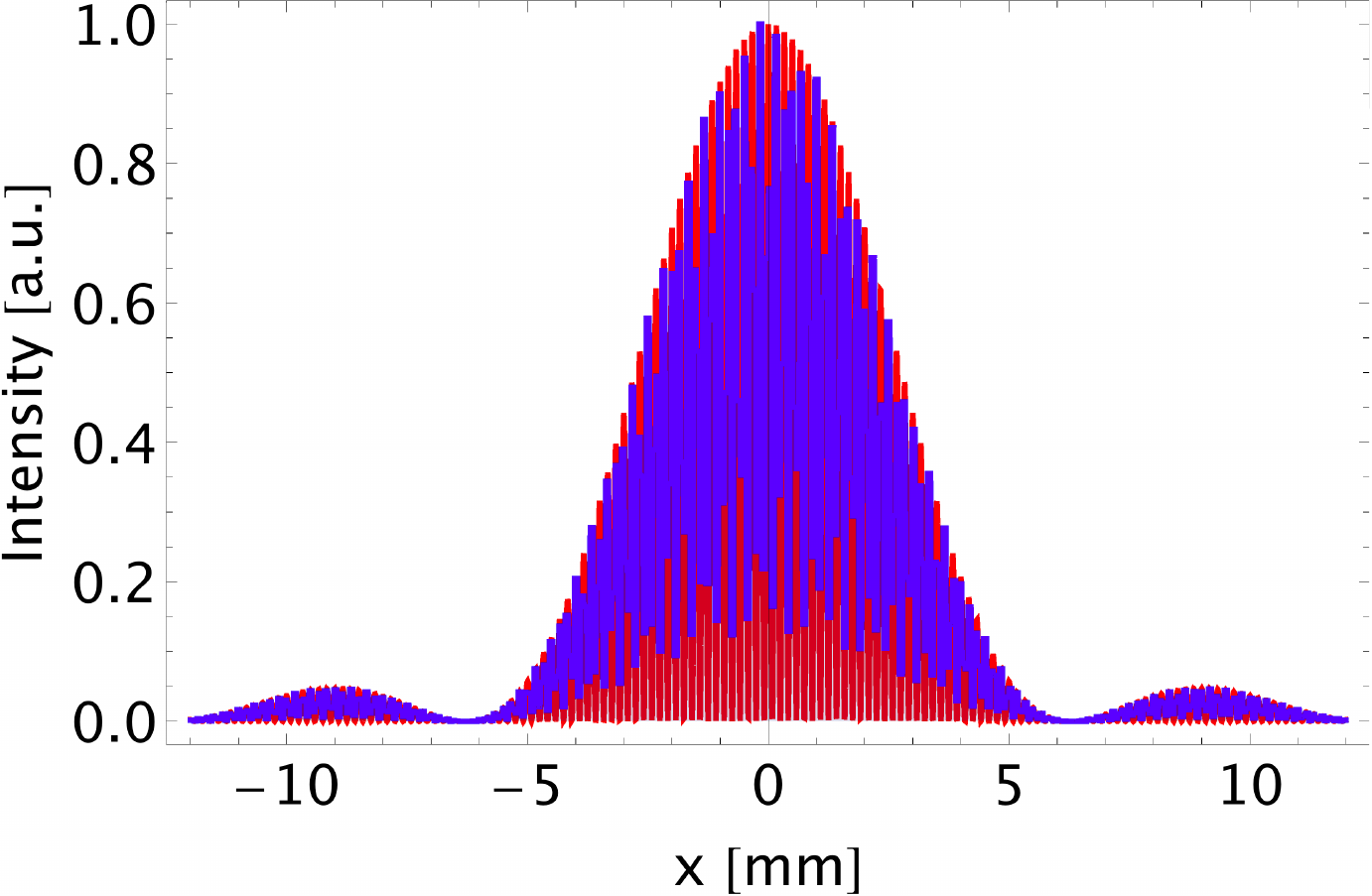}
\caption{(color online) Same as Fig. \ref{figura2} but for $\lambda=100$ nm. As the wavelength is decreasing, the fringes become denser with respect to the reference case. Thus, the detector starts to reproduce with poor fidelity the interference pattern and the output signal reflects more the envelope structure rather than the interference pattern. The fringe visibility in this case reduces to $V=0.644$.}
\label{figura3}
\end{center}
\end{figure}

\begin{figure}[t!]
\begin{center}
\includegraphics[width=0.5\textwidth]{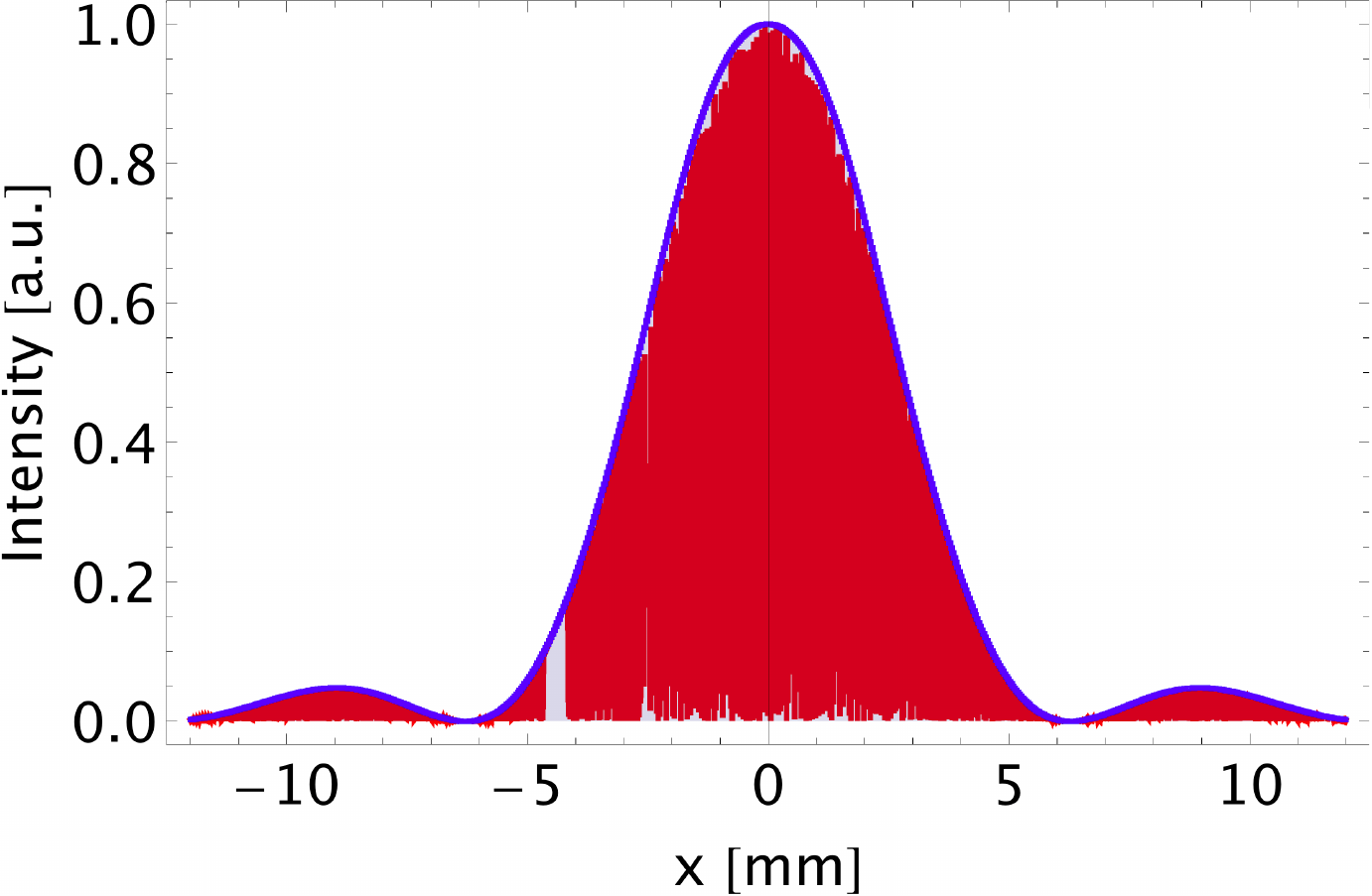}
\caption{(color online) Same as Fig. \ref{figura2} but for $\lambda=20$ nm. The fringe visibility for this case is $V=0$, i.e.,  in this case the fringes are so dense that the detector cannot reproduce its structure anymore but rather he detects the envelope of the interference pattern. This corresponds to the classical case for which we see no interference due to the fact that they are too tiny to be distinguished by any finite-size detector.}
\label{figura4}
\end{center}
\end{figure}

In Figs. \ref{figura2} to \ref{figura4} the analytic expression of the interference pattern is shown with a solid thick red line, together with the signal retrieved by the detectors with vertical blue histograms. As one can see, passing from Fig. \ref{figura2} (where $\lambda=800$ nm) to Fig. \ref{figura4} ($\lambda=20$ nm) the wavelength becomes smaller, the fringes on the screen become denser, and the detectors are not able anymore to distinguish between two adjacent fringes. In the last case (Fig. \ref{figura4}) they just generate, as output signal,  the average of a rapidly oscillating signal, i.e.,  its envelope.

The situation depicted in Fig. \ref{figura2} corresponds to the observation of a fringe pattern of a particle with tiny mass (or equivalently with large wavelength), i.e.,  the analogous of a \emph{quantum} particle, that undergoes diffraction from the double slit structure; as expected, fringes can be seen because the detector used to reveal them has the capability to detect with high fidelity the oscillations of the fringes. 

As the wavelength diminishes toward zero (and the mass of the correspondent particle grows), we are moving towards the classical limit, where the interference pattern on the screen is not visible anymore, and the image on the screen follows the rule of classical physics \cite{feynman}. In this limit (Fig. \ref{figura4}),  the interference pattern has still fringes, but (because of the very tiny wavelength) they are too close with respect to the detector local resolution. Thus, due to the averaging operation of the detector, the rapidly oscillating fringes will be mediated across the detector length, producing as output signal a no longer oscillating signal. The classical probability distribution of finding the \emph{classical} particle on the screen is then retrieved as expected \cite{nota}.

\section{IV. THE EFFECT OF PARTICLE DECOHERENCE}
In this section we want to discuss a different approach (with respect to the one presented in the previous section) for studying the transition from the quantum to the classical world. This alternative approach is based on considering a monochromatic field at a fixed wavelength (that we will assume to be $\lambda=800$ nm) and study the behavior of the interference pattern as a function of the coherence between the two different polarizations that the electromagnetic field will experience in passing through the two slits (see Fig. \ref{figura1}). We then want to show that the lack of coherence will bring to a destruction of the interference fringes in the same way as the growing of the particle's mass in the last section did.

\subsection{A. Randomization of the Polarization}

Let us now assume that the two polarizers $P1$ and $P2$ in Fig. \ref{figura1} are random polarizers. We describe those polarizers as unit vectors rotating into the $\{x,y\}$ plane with an angle $\theta_k$ ($k\in\{1,2\}$) being a random variable with a suitable well-behaved probability distribution $\rho(\theta_k)$ in $[0,2\pi]$.
The polarization unit vectors ar then given by $\hat{\mathbf{p}}_k=\cos\theta_k\hat{\mathbf{x}}+\sin\theta_k\hat{\mathbf{y}}$. 

In order to study the effect of decoherence on the interference fringes, we need to ideally perform a great number of experiments with different values of the two random variables 
$\theta_1$ and $\theta_2$ in order to span all the possible cases. Once we have a sufficient number of outcomes (that are grouped into an ensemble of measurements) we can perform on them  a statistical analysis in order to retrieve the needed informations. 

Although we can simulate this experiment with a numerical routine and easily obtain a large number of outcomes (each for any value of the random variable $\theta_k$) that we can later on analyze with some statistical tool, a more formal way to proceed (that allows us to retain the physical picture and the sense it bears with itself) is to introduce the so-called coherency matrix \cite{mandelWolf}
\begin{equation}
\mathbf{J}=
\begin{pmatrix}
  J_{11} &  J_{12}\\
  J_{21} &  J_{22}
 \end{pmatrix},
\end{equation}
whose elements $J_{ik}=\langle \mathbf{E}_i(x)^*\cdot\mathbf{E}_k(x)\rangle$ represent the average (over the whole statestical population of data acquired) of the scalar product between the two differently polarized fields, i.e., the \emph{coherences} between these two fields. The symbol $\langle\cdots\rangle$ means that this value has been already averaged over the whole ensemble (ensemble average). In this description, the electric field $\mathbf{E}_k(x)$, although formally can be still written as in Eq. \eqref{E_init}, becomes a stochastic field depending on the random variable $\theta_k$.

All the quantities related with the electric field have then to be treated as stochastic quantities too. In particular, if we want to calculate the total field intensity in the plane $z=L$ we can calculate it for the single element of the ensemble (for whom we can use Eq. \eqref{intensity}) and then average its result over the entire ensemble as foliows \cite{mandelWolf}:
\begin{equation}\label{stocastico}
\langle I(x,z)\rangle=\langle\mathbf{E}^*(x,z,t;\{\theta_1,\theta_2\})\cdot\mathbf{E}(x,z,t;\{\theta_1,\theta_2\})\rangle,
\end{equation}
where the dependence of the electric field from the two angular random variables has been explicitly written for stressing once more the stochastic nature of the field \cite{nota2}.

Substituting in Eq. \eqref{stocastico} the expression of the electric field as given by Eq. \eqref{E_init}, and using the definition of the elements of the coherency matrix, it is possible to rewrite the last equation in a more useful form as follows:
\begin{equation}\label{stoc2}
\langle I(x,z)\rangle=\Big[J_{11}+J_{22}+2J_{12}\cos\Big(\frac{kbx}{L}\Big)\Big].
\end{equation}
where $0\leq J_{12}=\cos(\theta_1-\theta_2)\equiv\delta\leq 1$. In obtaining the previous equation, we used the fact that the coherency matrix is Hermitian, thus $J_{ik}^*=J_{ki}$. 

As a first observation, we can say that if $\delta=0$ (namely if $\theta_1-\theta_2=n\pi/2$), the total intensity is simply given by the sum $J_{11}+J_{22}=|\mathbf{E}_1|^2+|\mathbf{E}_2|^2$, i.e., by the sum of  the intensity transmitted from each slit separately: $I_{tot}=I_1+I_2$. This is referred as the incoherent sum, as there is no correlation between the polarization of the field in one slit and the polarization in the other slit. In this case it is not difficult to see that, in order for the total intensity to be preserved, the two diffracted fields must have the same intensities(namely $J_{11}=J_{22}=1/2$). On the other hand, when $\delta=1$ we are in the case of full coherence, i.e., there is the maximum correlation between the polarizations of the two fields. This latter case corresponds (once the wavelength has been fixed) to the fringe pattern described by Eq. \eqref{intensity}. All the intermediate cases $0\leq\delta\leq 1$ correspond to situations of partial coherence between the two fields.

If we then normalize the result of Eq. \eqref{stoc2} to the total intensity (that we just calculated in the incoherent case to be $I_{tot}=J_{11}+J_{22}$), we obtain:
\begin{equation}
\langle I(x,L)\rangle=2\Big[1+\delta\cos\Big(\frac{kbx}{L}\Big)\Big].
\end{equation}
By artificially changing the value of $\delta$ we can then simulate the effect of polarization decoherence on the interference pattern generated by the two beams that diffracts from the apertures and interfere on the detector plane.

The aim of the next subsection is then to show that this decoherence gives the same results as the ones presented in the previous section, i.e.,  the classical behavior of a system (e.g. the Feynman's bullet) can be explained not only by saying that a classical system does not show quantum signatures due to its high mass (i.e.,  tiny wavelength), but it can be even explained by saying that the wavefunction associated to each atom that constitute the classical particle experiences decoherence while interacting with the other atom's wavefunction. This means that the quantum features of all of these atoms are averaged over the ensemble represented by the whole atoms themselves;
this average operation destroys any quantum effect, retrieving once again the classical probability distribution for the particle.

\subsection{B. Classical Limit for Decoherent Waves}
In this section we present the results of the two slit experiment, where now the two beams experiment two different polarizations when exiting from the left and right slit. The polarization acquired by the two fields is random, as described above. The geometrical parameters remains the same as for the previous case, but now the wavelength is kept fixed at its reference value of $\lambda=800$ nm. The detector on the screen is also modeled in the same way as it was in the previous section, and the effect of polarization randomization is accounted, as described before, with the help of the $\delta$ variable, that can take all the values inside the interval $[0,1]$.

\begin{figure}[t!]
\begin{center}
\includegraphics[width=0.5\textwidth]{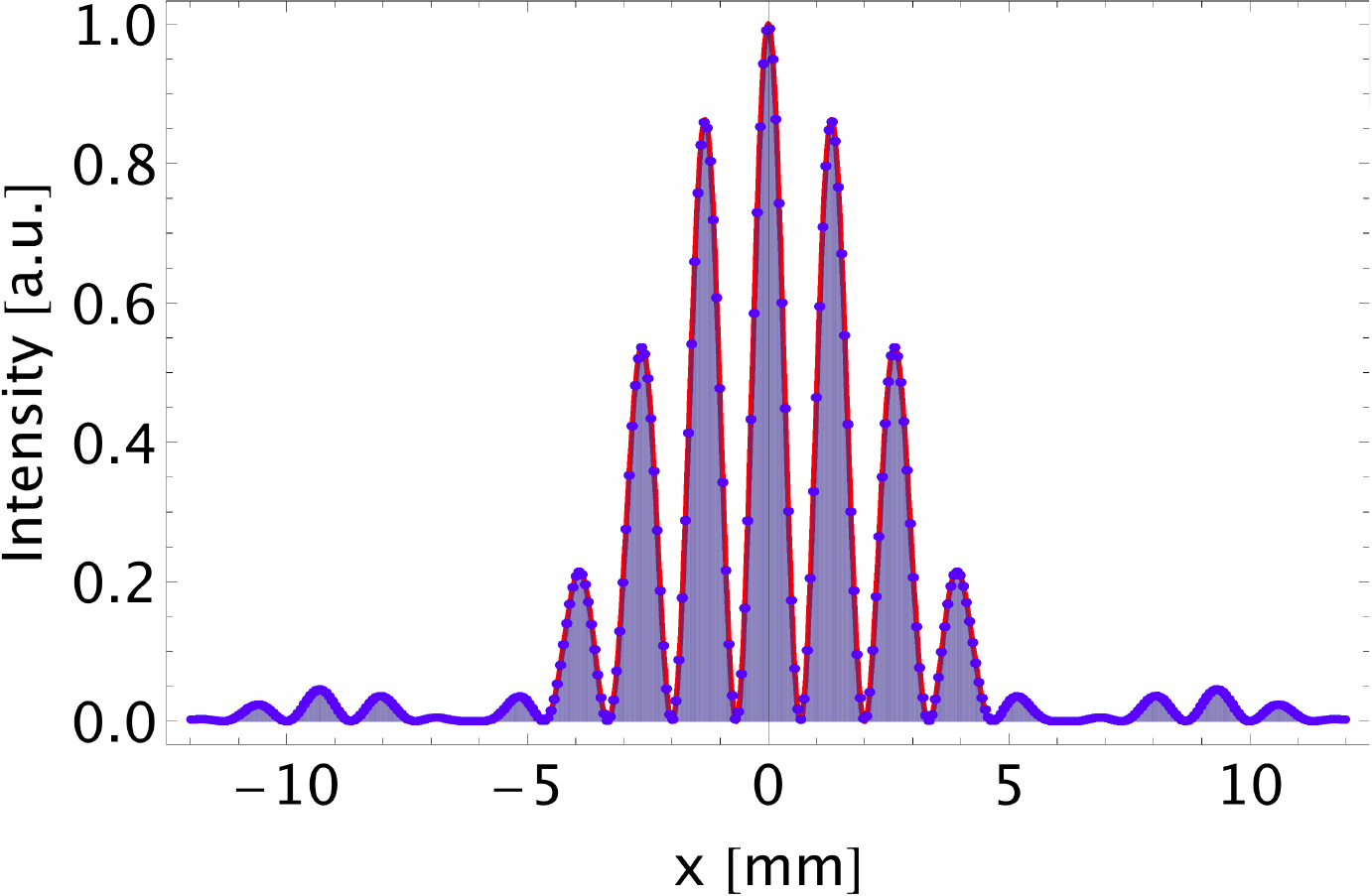}
\caption{(color online) Analytic expresison (solid red line) and detector output (blu vertical bars) for the interference pattern for the case $\delta=1$, i.e.,  complete coherence between the two beams exiting from the two slits. This case is exactly analogue to the one presented in Fig. \ref{figura2}.}
\label{figura5}
\end{center}
\end{figure}

\begin{figure}[t]
\begin{center}
\includegraphics[width=0.5\textwidth]{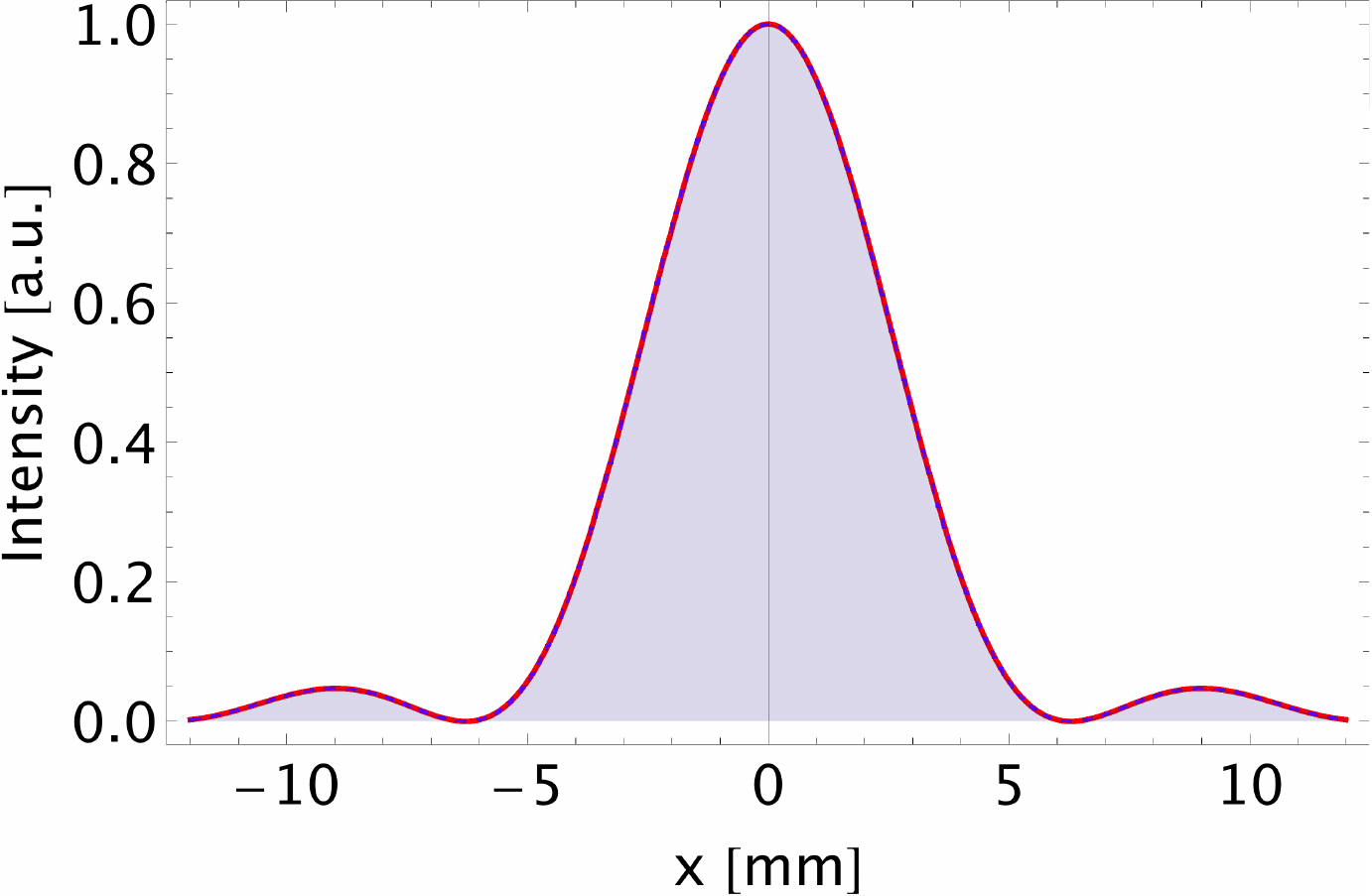}
\caption{(color online) Same as Fig. \ref{figura5} but for the case $\delta=0$, i.e.,  complete incoherence between the two beams exiting from the two slits. The analytical solution is shown in red, while the detected intensity is shown in blue dashed lines. Note that in this case the \emph{classical} probability distribution is retrieved.}
\label{figura6}
\end{center}
\end{figure}

\begin{figure}[!t]
\begin{center}
\includegraphics[width=0.5\textwidth]{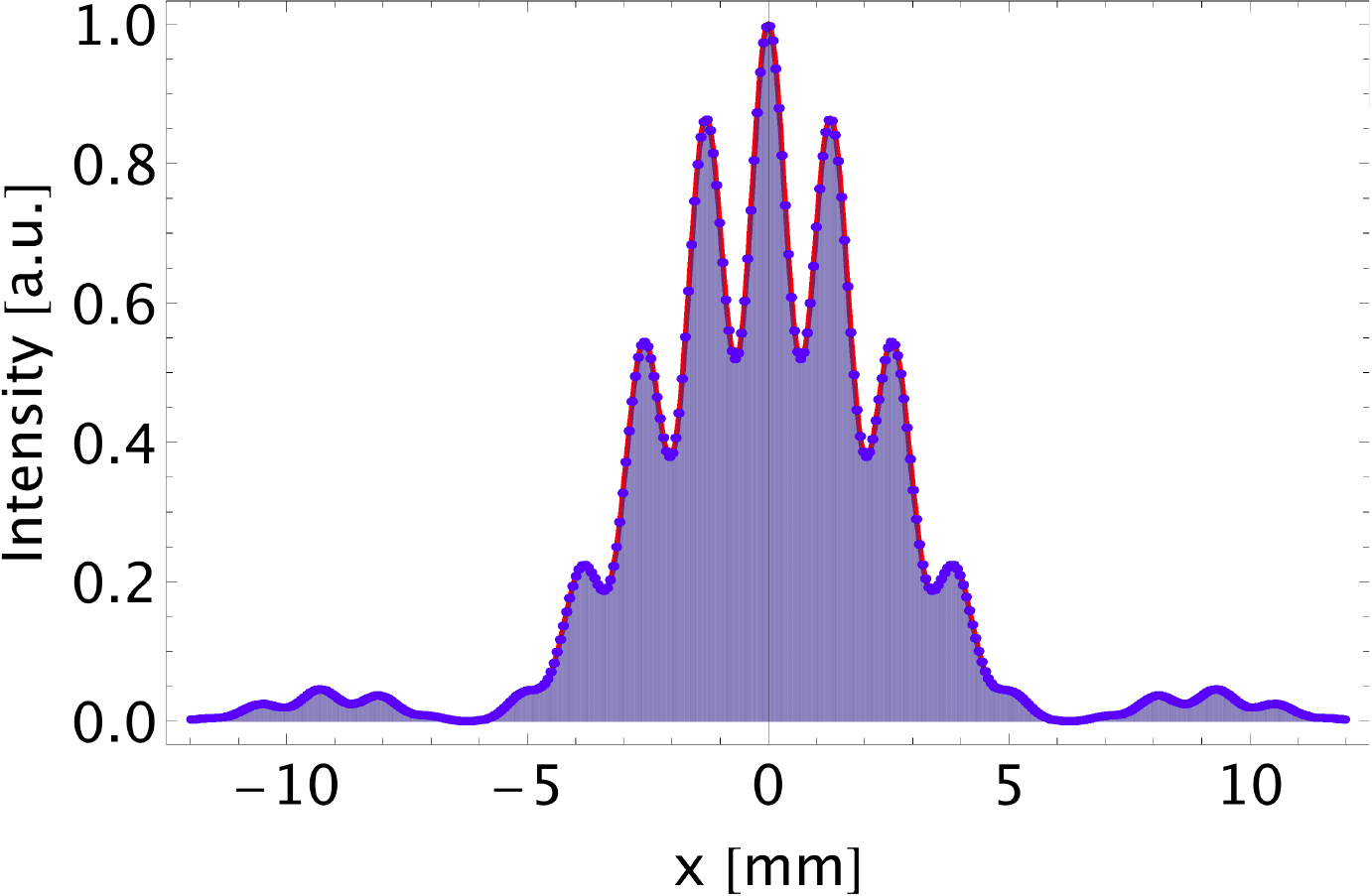}
\caption{(color online) Same as Fig. \ref{figura5} but for the case $\delta=0.3$. Note how in this case the fringes start to show up even if their visibility is very low, namely $V=0.314$. The partial coherence of the two beams, however, is sufficient to make the fringes to appear.}
\label{figura7}
\end{center}
\end{figure}

\begin{figure}[!t]
\begin{center}
\includegraphics[width=0.5\textwidth]{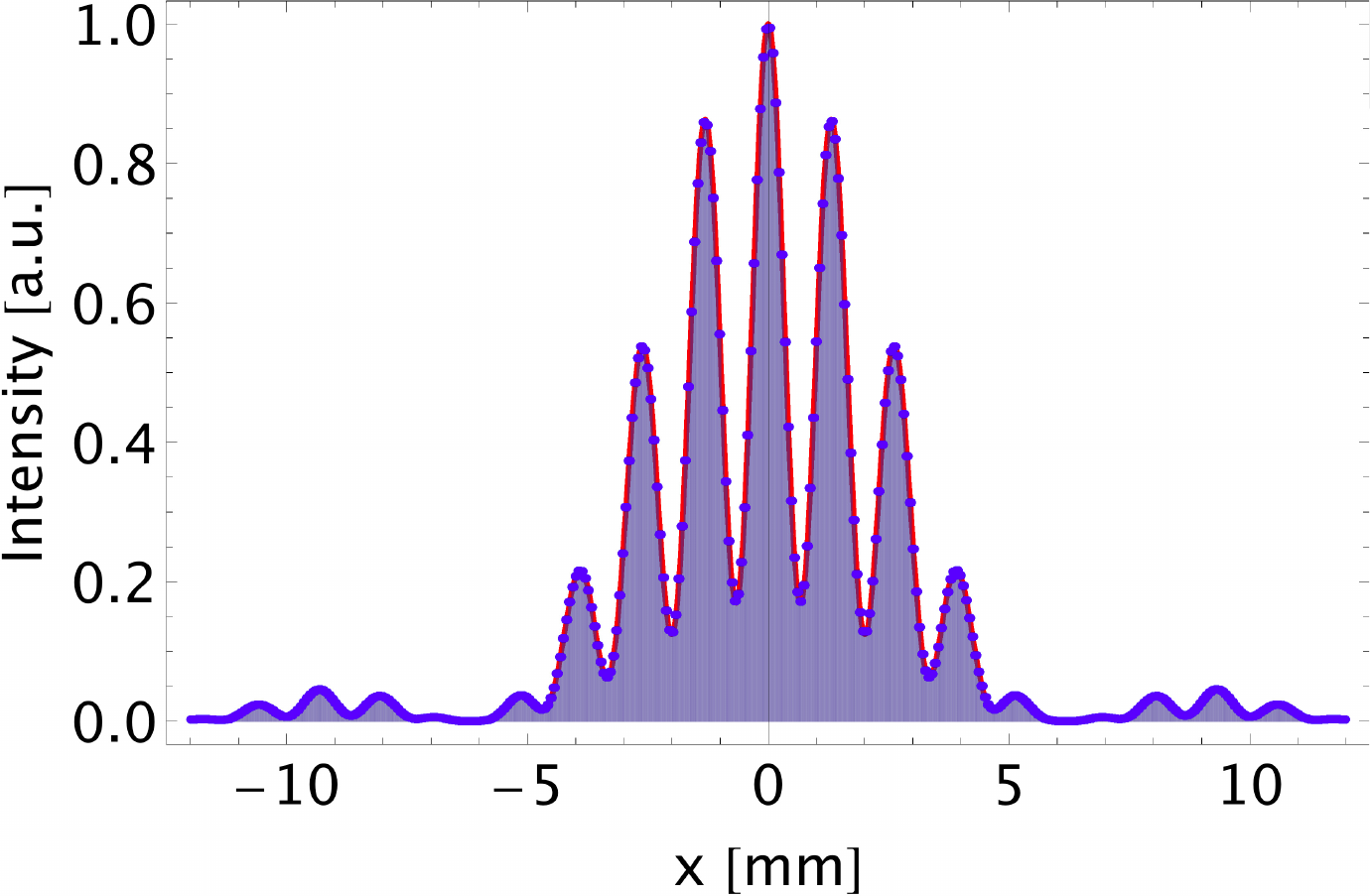}
\caption{(color online) Same as Fig. \ref{figura5} but for the case $\delta=0.7$. The fringe pattern in this case is more pronounced than Fig. \ref{figura7} due to the high visibility of the fringes, that in this case is $V=0.705$.}
\label{figura8}
\end{center}
\end{figure}

Figures \ref{figura5} to \ref{figura7} present either the analytic expression (the solid thick red line) and the detector output (the blue vertical histograms) that one can see on the screen while the value of the $\delta$ parameter is changed from the case of complete coherence of the two beams ($\delta=1$, Fig. \ref{figura5}) where interference takes place exactly as seen in Fig. \ref{figura2}, to the case of complete incoherence of the two beams ($\delta=0$, Fig. \ref{figura6}) when the two beams sum incoherently, giving  $I_{tot}=I_1+I_2$. All the intermediate cases (see e.g. Figs. \ref{figura7} and \ref{figura8}) witness how the partial coherence of the two beams affects the interference pattern, by mainly reducing the visibility of the fringes as the value of $\delta$ changes from 1 towards 0.\\
The situation depicted in Fig. \ref{figura6} is analogous to the case depicted in Fig. \ref{figura4}; in this way, it is possible to say that the absence of interference fringes for a classical object (i.e.,  the lack of observation of \emph{quantum} effects from it) can be either explained as the heavy mass carried by the classical object that makes its wavelength very tiny that no detector can distinguish it, \emph{or} it can be explained even by saying that the various wave functions associated to each of the atoms that constitute the object experiment decoherence; this is then responsible of incoherent sum, bringing them to a situation where no interference occurs.

Figures \ref{figura7} and \ref{figura8} also show that for the intermediate case of partial coherence between the two beams, some fringes arise, even if their visibility is not one. This situation is analogous, in principle, to the one depicted in Fig. \ref{figura3}, when the detector starts to not reproduce with high fidelity the fringe pattern. Even if in this case there is not this kind of problem (the wavelength is kept fixed, so the detector works all the time at its best performances), the partial coherence operates in exactly the same way, providing a fringe pattern that approaches the incoherent pattern ad the value of $\delta$ is moved towards zero.

\section{V. CONCLUSIONS}
In this work, we have given a simple yet insightful test of Feynman's idea on the two slit experiment, using two different approaches. In the first case, the classical limit was realized by accounting for the averaging effect of the detector appearing in the limit of short wavelength (compared to the detector's dimension). In the second case, instead, the classical limit was obtained by controlling the degree of spatial coherence of the impinging light field. Here, the classical limit corresponds to the case of a fully incoherent field. In both cases, we obtained the same result for the ``classical limit" namely the disappearance of interference fringes, thus vindicating and validating Feynman's belief.
%

%Bibliography


\begin{thebibliography}{99}
\bibitem{feynman}
R. P. Feynman, R. B. Leighton and M. Sands, \emph{The Feynman Lectures on Physics, Vol. III: Quantum Mechanics}, the new millennium edition (Basic Books, New York, NY,  2011). \\

\bibitem{goodman}
J. W. Goodman, \emph{Introduction to Fourier Optics}, second edition (McGraw Hill, New York, NY, 2000) .

\bibitem{pakstakos}
G. Patsakos, Am. J. Phys. \textbf{44}, 158(1976)
\bibitem{nota}
Note that the classical (i.e.,  incoherent) light distribution on the screen is simply given by the incoherent sum of light coming from the first slit when the second is closed and viceversa. In this case, however, the light pattern coming from \emph{only} the first or second slit has the same spatial distribution on the screen. Then, the incoherent sum light distribution would have the same shape as the single slit light distribution but with doubled intensity.
\bibitem{peres_dec}
A. Peres, \emph{Quantum theory: concepts and methods} (Springer, Berlin, 1995).
\bibitem{mandelWolf} L. Mandel and E. Wolf, \emph{Optical coherence and quantum optics} (Cambridge U. P., New York, NY, 1995).
\bibitem{nota2} Note also that while the electric field per se is a function of time, the total intensity is time independent, since the electric field (although stochastic) is assumed to be monochromatic. 

\end{thebibliography}
\end{document}